 \documentstyle[12pt]{article}
  
 \renewcommand{\title}[1]{\null\vspace{25mm}
   \noindent{\Large{\bf #1}}\vspace{10mm}
    }
 \newcommand{\authors}[1]{\noindent{\large #1}\vspace{20mm}
    }
 \newcommand{\address}[1]{{\center{\noindent #1\vspace{10mm}}
    }}
 \renewcommand{\abstract}[1]{\vspace{17mm}
  \noindent{\small{\em Abstract.} #1}\vspace{2mm}
   }     
 
 \newcommand{\en}{\begin{equation}} \newcommand{\dv}{\end{equation}} \newcommand{\de}{\partial}         
 \newcommand{\dms}{\partial_{\mu}}  
  
 \newcommand{\tr}{\begin{description}}
 \newcommand{\st}{\end{description}} 
 \newcommand{\pr}{\hspace{1mm}}     \newcommand{\ptr}{\hspace{3mm}}
 \newcommand{\pe}{\begin{eqnarray}} \newcommand{\se}{\end{eqnarray}}

 \newcommand{\ro}{\varrho}  
 \newcommand{\sed}{\begin{array}}  \newcommand{\os}{\end{array}}  
 \newcommand{\ee}{\varepsilon}
 \newcommand{\la}{\lambda} \newcommand{\si}{\sigma} \newcommand{\al}{\alpha}
 \newcommand{\da}{\delta}  
 \newcommand{\nm}{\nonumber}
  
 \newcommand{\pp}{\hspace{5mm}}\newcommand{\am}{A_\mu}

 \newcommand{\dm}{\da_\mu}

  \newcommand{\Tr}{\hbox{Tr}} 
 \newcommand{\bd}{\begin{Def}\hspace{-2mm}: \rm}

 \newcommand{\Sigmanull}{\Sigma^{(0)}}
 
\begin{document}   \setcounter{table}{0}
 
 \begin{titlepage}
 \begin{center}
 \hspace*{\fill}{{\normalsize \begin{tabular}{l}
                              {\sf REF. TUW 99-07}\\
                              \end{tabular}   }}

 \title{Generalized 2D BF Model quantized in the axial gauge}

 \authors{R. Leitgeb$^1$, J. Rant$^1$, M.~Schweda and H. Zerrouki}    \vspace{-20mm}
       
 \address{Institut f\"ur Theoretische Physik, 
                          Technische Universit\"at Wien\\
      Wiedner Hauptstra\ss e 8-10, A-1040 Wien, Austria}
      
 \footnotetext[1]{Work supported in part by the "Fonds zur F\"orderung der
 Wissenschaflicher Forschung", under Project Grant Number P11354-PHY.}
        
 \end{center} 
 \thispagestyle{empty}
 \abstract{We discuss the ultraviolet finiteness of the two-dimensional BF model 
 coupled to topological matter quantized in the axial gauge. This noncovariant gauge fixing
 avoids the infrared problem in the two-dimensional space-time. The BF model together
 with the matter coupling is obtained by dimensional reduction of the ordinary 
 three-dimensional BF model. This procedure furnishes the usual linear vector supersymmetry
 and an additional scalar supersymmetry. The whole symmetry content of the model allows
 to apply the standard algebraic renormalization procedure which we use to prove that 
 this model is ultraviolet finite and anomaly free to all orders of perturbation theory.
 }
 \end{titlepage}
 
 \section{Introduction}
 
 Topological field models have been an object of intense interest over the last decade and
 brought significant developments to the understanding of the topology and geometry of
 low dimensional manifolds \cite{Hor,Don}. The main property of the topological models
 \cite{Bir} is the fact that the observables depend only on the global structure of
 the space-time manifold on which they are defined, e.g. no physical degrees of freedom
 exist locally. In particular, they are independent of any
 metric which may be used to define the classical theory. 
 \\[2mm]
 There are two different types of topological field theories. The first one is called 
 Witten-type \cite{Witten} whose whole gauge fixed action may be written as a total
 BRS variation. The most prominent example of Witten-type models is the topological 
 Yang--Mills theory. The second type of topological models are the Schwarz-type models 
 characterized by the fact that only the gauge fixed part of the action is an exact
 BRS-variation. The examples 
 of the Schwarz-type models are Chern-Simons and BF theories. A common feature of such models
 is the presence of the so-called topological linear vector supersymmetry. 
 The corresponding operator $\da_\mu$ and the usual BRS-operator $s$ form a graded algebra 
 of Wess-Zumino type:
 $$ \{s,\da_\mu\}=\de_\mu \ptr. $$
 The aim of the present work is to analyze the ultraviolet behavior of two-dimensional BF model 
 with a matter coupling \cite{Cham} quantized in the axial gauge. 
 The infrared and ultraviolet behavior of the pure two-dimensional BF model has been 
 already discussed in \cite{Blasi,Bossino}. 
 Usually the propagators in two space-time dimensions are not
 well-defined in the infrared region. In the present work, however, the use of the axial
 gauge removes the singular behavior at long distances. To carry out the proof of 
 perturbative finiteness we will use the ordinary algebraic renarmalization procedure in
 the context of BRS-symmetry \cite{Becchi,Pig,Buch}. Besides the usual BRS-symmetry there
 exists also the topological linear vector supersymmetry \cite{Pig} together with
 an additional topological scalar supersymmetry. The latter scalar supersymmetry is a
 by-product in reducing the three-dimensional BF model to two-dimensional space-time. 
 The whole symmetry content of the model is basis for the proof of the ultraviolet finiteness.
 \\[2mm]
 The paper is organized as follows. In Section 2 we present the classical 
 algebraic properties of the model at the classical level. We display the BRS-transformations, 
 the topological linear vector supersymmetry and the additional scalar supersymmetry. In 
 Section 3 we discuss
 the proof of perturbative finiteness of the model by analyzing its stability. The same
 arguments as in \cite{Leitgeb} imply that the model is free of anomalies.
 
 \section{The classical model}
 
 The action of classical BF model living on manifolds ${\cal M}$ with $(n+2)$-dimensions can 
 be defined \cite{Hor,Bir,Pig} according to 
 \en S_{BF}=\Tr \int_{\cal M} BF={1\over 2n!}\Tr\int_{\cal M} d^{n+2}x\,\ee^{\mu_{1}\ldots
    \mu_{n+2}}B_{\mu_{1}\ldots\mu_{n}}F_{\mu_{n+1}\mu_{n+2}} \ptr, \dv
 where $F$ is a two-form 
 \en F=dA+{1\over 2}[A,A]={1\over 2}F_{\mu\nu}dx^\mu dx^\nu, \dv
 A is the usual gauge connection one-form and
 B is a n-form:
 \en B={1\over n!}B_{\mu_1\ldots\mu_n}dx^{\mu_1}\ldots dx^{\mu_n} \ptr. \dv
 This action has a topological character since it is 
 independent of the metric of the manifold ${\cal M}$. 
 \\[2mm]
 In the case of two-dimensional flat Euclidean space-time the action of the BF model reads 
 \en S_{inv}^{(1)}={1\over 2}\int_{\cal M}d^2x\,\ee^{\mu\nu}F_{\mu\nu}^a\phi^a\ptr, 
     \label{Sinv}   \dv
 where $\ee^{\mu\nu}$ is the totally antisymmetric Levi-Civita tensor (with $\ee^{12}=+1$),
 $\phi^a$ is a scalar field and $F_{\mu\nu}^a$ is the field strength given by
 \en F_{\mu\nu}^a=\dms A_\nu^a-\de_\nu\am^a+f^{abc}\am^bA_\nu^c \ptr. \dv
 Here, $\am^a$ is the gauge field with the group index $a$. All fields belong to the
 adjoint representation of some compact semi-simple gauge group $G$ whose structure
 constants $f^{abc}$ are completely antisymmetric in their indices. The generators of 
 the Lie algebra are chosen to be anti-hermitian and subject to $[T^a,T^b]=f^{abc}T^c$
 and $\Tr(T^aT^b)=\da^{ab}$. 
 \\[2mm]
 The action (\ref{Sinv}) is invariant under the following infinitesimal gauge symmetry
 \pe \da^{(1)}\am^a &=& \dms\theta^a+f^{abc}\am^b\theta^c\equiv (D_\mu\theta)^a \ptr, \nm\\
     \da^{(1)}\phi^a&=& -f^{abc}\theta^b\phi^c \ptr, 
     \se    
    where $\theta^a$ and $D_\mu$ stand for a local gauge parameter and the covariant 
 derivative respectively. This action has much been investigated in \cite{Blasi,Bossino}.  
 Following \cite{Cham,Leitgeb} we enlarge the model by adding to (\ref{Sinv}) the following
 topological matter interaction term:
 \en S_{inv}^{(2)}=\int_{\cal M}d^2x\,\ee^{\mu\nu}(D_\mu B_\nu)^aX^a \ptr, \label{Smat} 
     \dv
 with two additional fields: a vector field $B_\mu^a$ and a scalar field $X^a$.
 The action (\ref{Smat}) is invariant under $\da^{(1)}$:
 \pe \da^{(1)}B_\mu^a &=& -f^{abc}\theta^bB_\mu^c \ptr, \nm \\
     \da^{(1)}X^a &=& -f^{abc}\theta^bX^c \ptr. 
     \se
 The introduction of the matter coupling (\ref{Smat}) implies that the total action
 \en S_{inv}=S_{inv}^{(1)}+S_{inv}^{(2)} \dv
 possesses an additional gauge symmetry given by 
 \pe \da^{(2)}\am^a &=& 0 \ptr, \nm \\
     \da^{(2)}\phi^a &=& -f^{abc}\Theta^bX^c \ptr, \nm \\
     \da^{(2)} X^a &=& 0  \ptr, \nm \\
     \da^{(2)} B_\mu^a &=& (D_\mu\Theta)^a \ptr, 
     \se
 where $\Theta^a$ is another infinitesimal local parameter. 
 \\[2mm]
 As usual, the quantization of gauge field models requires a gauge fixing in order to
 guarantee the existance of the gauge field propagators. This is done consistently by the
 introduction of the Faddeev-Popov ghost fields in the context of the BRS-framework
 \cite{Becchi}. Since there are
 two different gauge symmetries in the model, the BRS-quantization procedure requires
 two sets of ghost fields with the corresponding Lagrange multiplier fields. We therefore
 introduce two Faddeev-Popov ghosts $(c^a,\la^a)$ with the corresponding antighosts ($\bar c^a,
 \bar\la^a)$ and two Lagrange multipliers $(b^a,d^a)$. We choose the axial gauge and add  
 the following gauge fixing term to the action $S_{inv}$:
 \pe S_{gf} &=& s\int_{\cal M}d^2x\,\left(\bar c^an^\mu\am^a+\bar\la^an^\mu B_\mu^a\right)
     \nm\ptr, \\
     &=& \int_{\cal M}d^2x\,\left(b^an^\mu\am^a+d^an^\mu B_\mu^a-\bar c^an^\mu(D_\mu c)^a-
     \bar\la^an^\mu(D_\mu\la)^a+f^{abc}\bar\la^ac^bn^\mu B_\mu^c\right),\hspace{10mm} \se
 where $n^\mu$ is a fixed gauge direction. 
 The complete gauge fixed action
 is, by construction, BRS-invariant
 \en s(S_{inv}+S_{gf})=0 \ptr, \dv
 where the nilpotent and nonlinear BRS-transformation read as
 \pe s\am^a &=& (D_\mu c)^a \ptr, \nm \\
     sB_\mu^a &=& (D_\mu\la)^a-f^{abc}c^bB_\mu^c \ptr, \nm \\
     s\phi^a &=& -f^{abc}c^b\phi^c-f^{abc}\la^bX^c \nm \ptr, \\
     sX^a &=& -f^{abc}c^bX^c\ptr, \nm \\
     sc^a &=& -{1\over 2}f^{abc}c^bc^c \nm \ptr, \\
     s\la^a &=& -f^{abc}c^b\la^c \nm \ptr, \\
     s\bar c^a &=& b^a \pr, \hspace{1cm} sb^a=0 \ptr, \nm \\
     s\bar\la^a &=& d^a \pr, \hspace{1cm} sd^a=0 \ptr, \nm \\
     s^2 &=&0 \ptr. \label{BRS} \se 
 In order to control the $n^\mu$-dependance of the theory one enlarges the 
 BRS-transformations  by allowing also a variation of the axial vector
 $n^\mu$ \cite{ndep2}:
 \en s¸n^\mu=\chi^\mu\pr,\pp s\chi^\mu=0\ptr, \label{BRSn} \dv
 and by adding the following term to the action
 \en S_n=-\Tr\int_{\cal M}d^2x\,\left(\bar c^a\chi^\mu\am^a+\bar\la^a\chi^\mu B_\mu^a
     \right) \ptr. \dv
 Here, $\chi^\mu$ is an anticommuting parameter. Obviously, the new action
 \en S=S_{inv}+S_{gf}+S_n \label{S}\dv
 is now invariant under the enlarged BRS-transformations (\ref{BRS}) together with 
 (\ref{BRSn}). The physical situation is represented by putting $\chi^\mu$ to zero. 
 We present the canonical dimensions and the Faddeev--Popov charges of all fields in 
 Table 1. 
 \begin{center}
 \begin{table}[h] \label{fields}
 \begin{center} 
 \begin{tabular}{|c|c|c|c|c|c|c|c|c|c|c|c|c|} 
 \hline
  & $\am^a$ & $B_\mu^a$ & $\phi^a$ & $X^a$ & $c^a$ & $\la^a$ & $\bar c^a$ & $\bar\la^a$ &
    $ b^a$ & $d^a$  & $n_\mu$ & $\chi^\mu$\\
 \hline
 dim & 1 & 1 & 0 & 0 & 0 & 0 & 1 & 1 & 1 & 1  & 0 & 0 \\
 \hline
 $\phi\pi$ & 0 & 0 & 0 & 0 & 1 & 1 & -1 & -1 & 0 & 0  & 0 & 1 \\
 \hline 
 \end{tabular} 
 \caption{Dimensions and Faddeev--Popov charges of the fields}
 \end{center}
 \end{table}
 \end{center}
 As usual for topological field models
 the action (\ref{S}) possesses besides the BRS-symmetry an invariance with respect to the 
 linear vector supersymmetry:
 \en
 \begin{tabular}{ll}
  $\dm A_\mu^a=0$\ptr, & $\dm B_\mu^a=0 $\ptr,  \\
  $\dm\phi^a=\ee_{\mu\nu}n^\nu\bar c^a\ptr, $& $\dm X^a=\ee_{\mu\nu}n^\nu\bar\la^a \ptr,$ \\
  $\dm c^a=\am^a \ptr,  $& $\dm\la^a=B_\mu^a \ptr, $ \\
  $\dm\bar c^a=0 \ptr,  $& $\dm\bar\la^a=0 \ptr, $ \\
  $\dm b^a=\dms\bar c^a$\ptr,  & $\dm d^a=\dms\bar\la^a \ptr,  $  \\
  $\dm n^\nu=0$\ptr, & $\dm\chi^\nu=0\ptr. $
 \end{tabular}   \label{vektorsusy}
 \dv
 One can easily verify that
 \en \dm S=0 \ptr. \label{dmS} \dv
 Moreover, the action (\ref{S}) is left invariant under a further topological 
 scalar supersymmetric transformation acting on the fields as follows
 \en
 \begin{tabular}{ll}
  $\da A_\mu^a=-\ee_{\mu\nu}n^\nu\bar\la^a$\ptr,&$\da B_\mu^a=-\ee_{\mu\nu}n^\nu\bar c^a$\ptr,\\
  $\da\phi^a=0 $& $\da X^a=0 \ptr,$ \\
  $\da c^a=X^a \ptr,  $& $\da\la^a=\phi^a \ptr, $ \\
  $\da\bar c^a=0 \ptr,  $& $\da\bar\la^a=0 \ptr, $ \\
  $\da b^a=0$\ptr,  & $\da d^a=0 \ptr,  $  \\
  $\da n^\mu=0$\ptr, & $\da\chi^\mu=0\ptr. $
 \end{tabular}   \label{neuesymm}
 \dv
 This symmetry\footnote{The use of this symmetry simplifies the the proof of the finiteness 
 of the theory as we will see later on.} 
 may be interpreted in the following manner.
 First note that two-dimensional BF model coupled to topological matter can be obtained by
 a dimensional reduction of the three-dimensional BF model. The symmetry (\ref{neuesymm})
 corresponds then to the third component of the topological vector supersymmetry of the 
 three-dimensional BF model \cite{Andy1}. 
 The BRS-operator (\ref{BRS}), the vector supersymmetry $\delta_\mu$ (\ref{vektorsusy}) and 
 the operator $\delta$ defined in (\ref{neuesymm}) form an 
 algebra of Wess-Zumino type which closes on-shell on the translations:
 \pe \{s,s\} &=& 0 \ptr, \nm \\
     \{s,\dm\}A_\nu^a &=& \dms A_\nu^a+\ee_{\mu\nu}{\da S\over \da\phi^a}\ptr, \nm \\
     \{s,\dm\}\phi^a &=& \dms\phi^a+\ee_{\mu\nu}{\da S\over \da A_\nu^a}\ptr, \nm \\
     \{s,\dm\}B_\nu^a &=& \dms B_\nu^a+\ee_{\mu\nu}{\da S\over \da X^a}\ptr, \nm \\
     \{s,\dm\}X^a &=& \dms X^a+\ee_{\mu\nu}{\da S\over \da B_\nu^a}\ptr, \nm \\
     \{s,\dm\}\psi^a &=& \dms\psi^a, \ptr\forall\psi^a\in\{c^a,\la^a,\bar c^a,\bar\la^a,b^a,
     d^a\}\ptr, \nm \\
     \{s,\da\}\am^a &=& -\ee_{\mu\nu}{\da S\over\da B_\nu^a}\ptr,  \nm \\
     \{s,\da\}B_\nu^a &=& -\ee_{\mu\nu}{\da S\over\da A_\nu^a}\ptr,  \nm \\
     \{s,\da\}\psi^a &=& 0, \ptr\forall\psi^a\in\{\phi^a,X^a,c^a,\la^a,\bar c^a,\bar\la^a,b^a,
     d^a\}\ptr. \label{onshell} \se
 Moreover, the following algebraic relations hold
 \pe
     \{\dm,\da_\nu\} &=& 0 \ptr,\nm \\ 
     \{\dm,\da\} &=& 0 \ptr, \nm \\
     \{\da,\da\} &=& 0 \ptr. \label{onshell2}\se
 In order to describe the BRS-symmetry content consistently at the functional level,
 we introduce a set of external sources coupled
 to the nonlinear BRS-variations of the quantum fields:
 \en S_{ext}=\int_{\cal M}d^2x\,\left[\Omega^{\mu a}(s\am^a)+L^a(sc^a)+\ro^a(s\phi^a)+
     \si^{\mu a}(sB_\mu^a)+D^a(s\la^a)+Y^a(sX^a)\right] \ptr. \dv
 We display the canonical dimensions and the Faddeev--Popov charges of the external sources
 in Table \ref{sources}.
 \begin{center}
 \begin{table}[ht] \label{sources} \begin{center} 
 \begin{tabular}{|c|c|c|c|c|c|c|} 
 \hline
  & $\Omega^{\mu a}$ & $L^a$ & $\ro^a$ & $\si^{\mu a}$ & $D^a$ & $Y^a$ \\
 \hline
 dim & 1 & 2 & 2 & 1 & 2 & 2  \\
 \hline
 $\phi\pi$ & -1 & -2 & -1 & -1 & -2 & -1  \\
 \hline 
 \end{tabular} 
 \caption{Dimensions and Faddeev--Popov charges of the external sources}
 \end{center}
 \end{table}
 \end{center}
 The complete action
 \en \Sigmanull=S_{inv}+S_{gf}+S_n+S_{ext} \label{Sigma} \dv
 obeys the Slavnov identity:
 \pe {\cal S}(\Sigmanull) &=&
     \int_{\cal M}d^2x\,\left[{\da\Sigmanull\over\da\Omega^{\mu a}}{\da\Sigmanull\over\da\am^a}+
     {\da\Sigmanull\over\da L^a}{\da\Sigmanull\over\da c^a}+{\da\Sigmanull\over\da\ro^a}{\da
     \Sigmanull\over
     \da\phi^a}+{\da\Sigmanull\over\da\si^{\mu a}}{\da\Sigmanull\over\da B_\mu^a}+
     \nm \right.\\  &+& \left.{\da\Sigmanull\over
     \da D^a}{\da\Sigmanull\over\da\la^a}+ 
     {\da\Sigmanull\over\da Y^a}{\da\Sigmanull\over\da X^a}+b^a{\da\Sigmanull\over\da\bar c^a}
     + d^a{\da\Sigmanull\over\da\bar\la^a}\right]+\chi^\mu{\de\Sigmanull\over\de n^\mu} =0 
     \ptr.\hspace{5mm}   \label{Slavnov} \se 
 The introduction of external sources induces a modified Ward-operator for the linear 
 vector supersymmetry ${\cal W}_\mu$: 
 \pe {\cal W}_\mu &=& \int_{\cal M}d^2x\left[\ee_{\mu\nu}\ro^a{\da\over\da A_\nu^a}+
     \ee_{\mu\nu}(n^\nu\bar c^a-\Omega^{\nu a}){\da\over\da\phi^a}+\am^a{\da\over\da c^a}+
     \dms\bar c^a {\da\over\da b^a}+L^a{\da\over\da\Omega^{\mu a}}+ \nm \right.\\  
     &+& \left.\ee_{\mu\nu}Y^a{\da\over\da B_\nu^a}+\ee_{\mu\nu}(n^\nu\bar\la^a-\si^{\nu a})
     {\da\over\da X^a}+B_\mu{\da\over\da\la^a}+\dms\bar\la^a{\da\over\da d^a}+D^a
     {\da\over\da\si^{\mu a}} \right] \ptr, \se
 and the vector supersymmetry is broken linearly
 \en {\cal W}_\mu\Sigmanull=\Delta_\mu \ptr, \label{Wmu}\dv
 where
 \pe \Delta_\mu &=& \int_{\cal M}d^2x\left[-\ro^a\dms\phi^a+\ee_{\mu\nu}n^\nu\ro^ab^a+
     \ee_{\mu\nu}n^\nu Y^ad^a-Y^a\dms X^a-\si^{\nu a}\dms B_\nu^a+ \nm \right. \\
     &+& \left.D^a\dms\la^a+
     L^a\dms c^a-\Omega^{\nu a}\dms A_\nu^a\right] \ptr. \label{breaking}\se
 Note that the breaking  (\ref{breaking}) is linear in the quantum fields and therefore
 harmless at the quantum level. Moreover, the topological scalar supersymmetry  
 is expressed by the following Ward-operator ${\cal D}$:
 \pe {\cal D}&=&\int_{\cal M}d^2x\left[\ee_{\mu\nu}(\si^{\nu a}-n^\nu\bar\la^a){\da\over\da
     \am^a}+\ee_{\mu\nu}(\Omega^{\nu a}-n^\nu\bar c^a){\da\over\da B_\mu^a}+X^a{\da\over\da
     c^a}+\right.\nm \\  &+&
     \left.\phi^a{\da\over\da\la^a}+D^a{\da\over\da \ro^a}+L^a{\da\over\da Y^a}\right]\ptr, \se
 with
 \en {\cal D}\Sigmanull=\Delta \ptr, \label{D}\dv
 where the breaking
 \en \Delta=\int_{\cal M}d^2x\,\ee_{\mu\nu}n^\mu(\si^{\nu a}b^a+\Omega^{\nu a}d^a)  \dv
 is linear in quantum fields, hence being harmless at the quantum level. 
 For later use we introduce the linearized Slavnov operator ${\cal S}_\Sigmanull$:
 \pe {\cal S}_{\Sigmanull} &=& 
     \int_{\cal M}d^2x\,\left[{\da\Sigmanull\over\da\Omega^{\mu a}}{\da\over\da\am^a}+
     {\da\Sigmanull\over\da \am^a}{\da\over\da\Omega^{\mu a}}+
     {\da\Sigmanull\over\da L^a}{\da\over\da c^a}+{\da\Sigmanull\over\da c^a}{\da\over\da L^a}+
     {\da\Sigmanull\over\da\ro^a}{\da\over\da\phi^a}
     +\nm \right. \\ &+& \left.{\da\Sigmanull\over\da \phi^a}{\da\over\da\ro^a}+
     {\da\Sigmanull\over\da\si^{\mu a}}{\da\over\da B_\mu^a}+{\da\Sigmanull\over\da B_\mu^a}
     {\da\over
     \da\si^{\mu a}}+{\da\Sigmanull\over\da D^a}{\da\over\da\la^a}+{\da\Sigmanull\over\da\la^a}
     {\da\over\da D^a}+\nm \right. \\ &+& 
     \left.{\da\Sigmanull\over\da Y^a}{\da\over\da X^a}+{\da\Sigmanull\over\da X^a}
     {\da\over\da Y^a}+ 
     b^a{\da\over\da\bar c^a}
     + d^a{\da\over\da\bar\la^a} \right] +\chi^\mu{\de\over\de n^\mu} \ptr. \se 
  The algebraic relations (\ref{onshell}) and (\ref{onshell2}) may be rewritten in terms
  of functional operators as follows:  
  \pe \{{\cal S}_{\Sigmanull},{\cal S}_{\Sigmanull}\} &=& 0 \nm \ptr, \\
      \{{\cal W}_\mu,{\cal W}_\nu\} &=& 0 \nm \ptr, \\
      \{{\cal D},{\cal D}\} &=& 0 \nm \ptr, \\
      \{{\cal S}_{\Sigmanull},{\cal D}\} &=& 0 \nm \ptr, \\
      \{{\cal D},{\cal W}_\mu\} &=& 0 \nm \ptr, \\
      \{{\cal W}_\mu,{\cal S}_{\Sigmanull}\} &=& {\cal P}_\mu \ptr, \se
  closing off-shell.
  Here, ${\cal P}_\mu$ is the Ward operator for translations
  \en {\cal P}_\mu=\int_{\cal M}d^2x\sum_{\Phi_i}\dms\Phi_i^a{\da\over\da\Phi_i^a} \ptr, 
      \label{P}\dv
  where $\Phi_i^a$ stands for all fields. 
  \\[2mm]
  At the classical level the total action (\ref{Sigma}) is now constrained 
  by the following functional identities:
  \begin{itemize}
  \item \underline{Gauge conditions}:
  \pe {\da\Sigmanull\over \da b^a} &=& n^\mu\am^a \ptr, \nm \\
      {\da\Sigmanull\over \da d^a} &=& n^\mu B_\mu^a \ptr. \label{Eichbed} \se
  \item \underline{Integrated ghost equations}:
  \pe {\cal F}^a\Sigmanull &=& \Delta_{\cal F}^a \ptr, \nm \\
      {\cal G}^a\Sigmanull &=& \Delta_{\cal G}^a \ptr, \label{ghosteq} \se
  where
  \pe {\cal F}^a &=& \int_{\cal M}d^2x\left({\da\over\da c^a}-f^{abc}\bar c^b{\da\over\da b^c}
      -f^{abc}\bar\la^b{\da\over\da d^c} \right) \ptr, \nm \\
      {\cal G}^a &=& \int_{\cal M}d^2x\left({\da\over\da\la^a}-f^{abc}\bar\la^b
      {\da\over\da b^c}\right)\ptr, \se
  and
  \pe \Delta_{\cal F}^a &=& \int_{\cal M}d^2x\,f^{abc}\left(-\ro^b\phi^c-Y^bX^c+L^bc^c+D^b\la^c
      -\Omega^{\mu b}\am^c-\si^{\mu b}B_\mu^c\right) \ptr, \nm \\
      \Delta_{\cal G}^a &=& \int_{\cal M}d^2x\,f^{abc}\left(D^bc^c-\ro^bX^c-\si^{\mu b}\am^c
      \right) \ptr. \se
  Note that, once again, the classical breakings $\Delta_{\cal F}^a$ and $\Delta_{\cal G}^a$
  are linear in the quantum fields. 
  \item \underline{Local antighost equations} which are obtained by
  commuting the gauge conditions with the Slavnov identity:
  \pe \bar{\cal F}^a\Sigmanull=\left({\da\over\da\bar c^a}+n^\mu{\da\over\da\Omega^{\mu a}}
      \right)\Sigmanull&=& 0 \ptr, \nm \\
      \bar{\cal G}^a\Sigmanull=\left({\da\over\da\bar\la^a}+n^\mu{\da\over\da\si^{
      \mu a}}\right)\Sigmanull &=& 0 \ptr. \label{anti}\se
 \item \underline{Ward identities of the rigid gauge invariance} which are obtained by
 commuting the ghost equations with the Slavnov identity: 
 \en {\cal H}^a\Sigmanull=\int_{\cal M}d^2x\sum_{\Phi_i}f^{abc}\Phi_i^b{\da\Sigmanull
     \over\da\Phi_i^c}=0\ptr, \label{H} \dv
 and
 \pe {\cal N}^a\Sigmanull &=& \int_{\cal M}d^2x\,f^{abc}\left(\am^b{\da\Sigmanull\over\da 
     B_\mu^c}+
     \si^{\mu b}{\da\Sigmanull\over\da\Omega^{\mu c}}+X^b{\da\Sigmanull\over\da\phi^c}+\ro^b
     {\da\Sigmanull\over\da  Y^c}+\right. \nm \\ &+& \left.c^b{\da\Sigmanull\over\da\la^c}+
     D^b{\da\Sigmanull
     \over\da L^c}+  \bar\la^b{\da\Sigmanull\over\da\bar c^c}+d^b{\da\Sigmanull\over\da b^c} 
     \right)=0 \ptr. \label{N} \se
 In (\ref{H}) $\Phi_i$ stands collectively for all fields.
 \end{itemize}

\section{Proof of the finiteness}

 This section is devoted to discuss the full symmetry content of the theory at the 
 quantum level, e.g. the question of possible anomalies and the stability problem
 which ammounts to analyze all invariant counterterms.  
 \\[2mm]
 We begin by studying the stability.
 This requires the analysis of the most general counterterms for the total
 action and implies to consider the following perturbed action
 \en \Sigma'=\Sigmanull+\Delta \ptr, \dv
 where $\Sigmanull$ is the total action (\ref{Sigma}) and $\Sigma'$ is a
 functional depending via $\Delta$ 
 on the same fields as $\Sigmanull$ and satisfying the Slavnov identity 
 (\ref{Slavnov}), the Ward identity for the
 vector supersymmetry (\ref{Wmu}), the two gauge conditions (\ref{Eichbed}), the two
 ghost equations (\ref{ghosteq}), the two antighost equations (\ref{anti}), the two 
 Ward identities of the rigid gauge invariance (\ref{H}) and (\ref{N}), the Ward
 identity for the translations (\ref{P}) as well as the Ward identity (\ref{D}). The 
 perturbation $\Delta$ collecting all appropriate invariant counterterms is an intergrated
 local field polynomial of dimension two and ghost number zero. 
 \\[2mm]
 Now we are searching for the most general deformation of the classical action such
 that the perturbed action $\Sigma'$ still fullfills the above constraints. The perturbation
 $\Delta$ must therefore obey the following set of equations:
 \pe {\da\Delta\over\da b^a} &=& 0 \ptr,  \label{gc1} \\
     {\da\Delta\over\da d^a} &=& 0 \ptr,  \label{gc2} \\
     {\da\Delta\over\da\bar c^a}+n^\mu{\da\Delta\over\da\Omega^{\mu a}}&=&0\ptr,\label{a1}\\   
     {\da\Delta\over\da\bar\la^a}+n^\mu{\da\Delta\over\da\si^{\mu a}}&=&0\ptr, \label{a2}\\ 
     {\cal S}_{\Sigmanull}\Delta &=& 0 \ptr,  \label{con1}\\
     {\cal W}_\mu\Delta &=& 0 \ptr,  \\
     {\cal P}_\mu\Delta &=& 0 \ptr,\\
     {\cal D}\Delta &=& 0 \ptr, \label{con2} \\
     \int_{\cal M}d^2x {\da\Delta\over\da c^a} &=&0 \ptr,  \label{ghostpert1}\\
     \int_{\cal M}d^2x {\da\Delta\over\da\la^a} &=&0 \ptr,  \label{ghostpert2}\\
     {\cal H}^a\Delta &=& 0 \ptr, \\
     {\cal N}^a\Delta &=& 0 \ptr. \se
 The first two equations (\ref{gc1}) and (\ref{gc2}) imply that the quantity $\Delta$
 does not depend on the multiplier fields $b^a$ and $d^a$. The validity of (\ref{a1}) and
 (\ref{a2}) implies that dependance of $(\Omega^{\mu a},\bar c^a)$ and $(\si^{\mu a},\bar\la^a)$
 is given by the following combinations
 \pe \tilde\Omega^{\mu a}&=&\Omega^{\mu a}-n^\mu\bar c^a \ptr, \nm \\
     \tilde\si^{\mu a} &=&\si^{\mu a}-n^\mu\bar\la^a \ptr. \se
 The equations (\ref{con1})--(\ref{con2}), as in reference \cite{Becchi2}, can be 
 collected into a unified operator $\da$:
 \en \da={\cal S}_{\Sigmanull}+\xi^\mu{\cal W}_\mu+\ee^\mu{\cal P}_\mu+\eta{\cal D}-
     \int_{\cal M}d^2x
     \xi^\mu{\de\over\de\ee^\mu}-\int_{\cal M}d^2x \zeta{\de\over\de\eta}\dv
 producing a single cohomology problem
 \en \da\Delta=0 \ptr. \label{cohproblem} \dv
 Here, $\xi^\mu$ and $\ee^\mu$ are constant vectors of ghost numbers +2 and
 +1 respectively and the quantities $\zeta$ and $\eta$ are constant scalars of ghost numbers
 +2 and +1 respectively. It can be easily verified that the operator $\delta$ is nilpotent
 \en \da^2=0 \ptr. \dv
 Due to the nilpotency of $\da$ any expression  of the form
 $\da\hat\Delta$ is automatically a solution of (\ref{cohproblem}). A solution of 
 this type is called a trivial solution. 
 Hence, the most general solution of (\ref{cohproblem}) reads 
 \en \Delta=\Delta_c+\da\hat\Delta \ptr. \dv
 Here, the nontrivial solution $\Delta_c$ is $\da$-closed $(\da\Delta_c=0)$, but not trivial 
 $(\Delta_c\ne\da\hat\Delta)$. Let us begin with 
 the determination of the nontrivial solution of (\ref{cohproblem}). For this purpose
 we introduce a filtering operator ${\cal N}$:
 \en {\cal N}=\int_{\cal M}d^2x\,\sum_{\Psi}\Psi{\da\over\da\Psi} \ptr, \dv
 where $\Psi$ stands for all fields, including $n^\mu,\chi^\mu,\xi^\mu$,$\ee^\mu,\zeta$
 and $\eta$.  
 To all fields we assign the homogeneity degree 1. 
 The filtering operator induces a decomposition
 of $\da$ and $\Delta$ according to
 \en \da=\da_0+\da_1+\ldots\pr,\pp \Delta=\Delta_1+\Delta_2+\ldots \label{decomp}\dv
 The operator $\da_0$ does not increase the homogeneity degree while acting on a field
 polynomial. On the other hand, the operator $\da_n$ increases the homogeneity degree by
 $n$ units. Similarly, $\Delta_n$ is a field polynomial of homogeneity degree $n$.
 Furthermore, the nilpotency of $\da$ leads now to 
 \en \da_0^2=0\pr,\pp \{\da_0,\da_1\}=0 \ptr. \dv
 Hence, we obtain from $\da\Delta=0$ the following relation
 \en \da_0\Delta_1=0\ptr, \dv
 with
 \en \Delta_1=\Delta_c^1+\da_0\hat\Delta_1\ptr.\dv
 The operator $\da_0$ reads:
 \en
 \begin{tabular}{ll}
  $\da_0 A_\mu^a=\dms c^a \ptr,$ & $\da_0 B_\mu^a=\dms\la^a$\ptr,\\
  $\da_0\phi^a=0 $& $\da_0 X^a=0 \ptr,$ \\
  $\da_0 c^a=0 \ptr,  $& $\da_0\la^a= \ptr, $ \\
  $\da_0 L^a=\dms\tilde\Omega^{\mu a}$\ptr,  & $\da_0 D^a=\dms\tilde\si^{\mu a} \ptr,  $  \\
  $\da_0\ro^a=\ee^{\mu\nu}\dms A_\nu^a\ptr,$ & $\da_0 Y^a=\dms A_\nu^a\ptr,$ \\
  $\da_0\tilde\Omega^{\mu a}=\ee^{\mu\nu}\de_\nu\phi^a$\ptr, & $\da_0\tilde\si^{\mu a}=
  \ee^{\mu\nu}\de_\nu X^a$ \ptr, \\
  $\da_0 n^\mu=\chi^\mu$\ptr, & $\da_0\chi^\mu=0\ptr, $  \\
  $\da_0\ee^{\mu}=-\xi^\mu \ptr,$ & $\da_0\xi^\mu=0 \ptr, $  \\
  $\da_0\zeta=-\eta\ptr, $  & $\da_0\eta=0$ \ptr. 
 \end{tabular}   \label{delta0}
 \dv
 We notice that the fields $n^\mu,\chi^\mu,\xi^\mu$,$\ee^\mu,\zeta$ and $\eta$
 transform under $\da_0$
 as doublets, being therefore out of the cohomology \cite {Brandt}. 
 The nontrivial solution
 $\Delta_c^1$ can now be written as intergrated local field polynomial of form degree two and
 ghost number zero:
 \en \Delta_c^1=\int_{\cal M}\omega^0_2 \ptr, \dv
 where $\omega^p_q$ is a field polynomial of form degree $q$ and ghost number $p$. 
 Using the Stoke's theorem, the Poincar\'e lemma \cite{Brandt} and the relation 
 $\{\da_0,d\}=0$, where $d$ represents the nilpotent exterior derivative ($d^2=0$),
 we obtain the following tower of descent equations:
 \pe \da_0\omega^0_2+d\omega_1^1 &=& 0 \ptr, \nm \\
     \da_0\omega_1^1+d\omega^2_0 &=& 0 \ptr, \nm \\
     \da_0\omega^2_0 &=&0 \ptr. \label{tower} \se
 The tower of descent equations (\ref{tower}) has been solved in \cite{Leitgeb}, where it
 was shown that the ghost equations (\ref{ghostpert1}) and (\ref{ghostpert2}) imply that the
 solution $\Delta_c^1$ must vanish identically. The usefulness of the decomposition  
 (\ref{decomp}) relies on a very general theorem stating that the cohomology of the
 complete operator $\da$ is isomorphic to a subspace of the cohomology of the operator $\da_0$. 
 \\[2mm]
 Next, we move to the computation of the trivial counterterms. These are constrained by the
 dimension and ghost number requirements. The scalar fields $\phi$ and $X$ both have 
 vanishing dimension and 
 ghost number zero, so that an arbitrary combination of them may appear infinitely many times
 in the counterterm. For the most general and possible combination of these fields we 
 use the notation $f^\al[\phi,X]$ as introduced in \cite{Leitgeb}:   
 \begin{equation}
 \label{falpha}
 f^\alpha[\phi,X]=\sum_{\{n_i\},\{m_i\}=0}^\infty \beta^\alpha_{n_i,m_i}\Bigg(
 \prod_{i=0}^\infty
 \phi^{n_{i}} X^{m_{i}}\Bigg)\,,
 \end{equation}
 where $\{n_i\}$ and $\{m_i\}$ are understood as $\{n_0,n_1,\dots\}$ and
 $\{m_0,m_1,\dots\}$, respectively.  Here, 
 $\beta^\alpha_{n_i,m_i}$ are constant coefficients to be determined.
 The most general trivial counterterm $\delta \hat{\Delta}$ where $\hat{\Delta}$
 has dimension 2 and carries ghost number -1 reads:
 \begin{eqnarray}\label{counter}
 \delta \hat{\Delta} &=& \delta \int_{\cal M} d^2 x \hbox{Tr} \left(\frac{}{}
 \ro f^1 +  Y f^2 +
 \tilde{\Omega}^\nu f^3 A_\nu f^4+
 \varepsilon_{\mu\nu}\tilde{\Omega}^\mu f^5 A^\nu f^6+
 n_\mu n_\nu \tilde{\Omega}^\mu f^7 A^\nu f^8 +
 \right.\nonumber\\
 &+&
 \varepsilon^{\mu\nu}n_\mu  \tilde{\Omega}_\nu f^9 n^\rho A_\rho
 f^{10}+
 n^\mu \tilde{\Omega}_\mu f^{11} \varepsilon^{\nu\rho}n_\nu A_\rho
 f^{12}+
 \varepsilon^{\mu\nu}n_\mu \tilde{\Omega}_\nu f^{13}
 \varepsilon^{\rho\si}n_\rho A_\si f^{14}+
 \nonumber\\
 &+&
 \tilde{\si}^\nu f^{15} A_\nu f^{16} +
 \varepsilon_{\mu\nu}\tilde{\si}^\mu f^{17} A^\nu f^{18} +
 n_\mu n_\nu \tilde{\si}^\mu f^{19} A^\nu f^{20}+
 \nonumber\\
 &+&
 n^\mu \tilde{\si}_\mu f^{23} \varepsilon^{\nu\rho}n_\nu A_\rho f^{24}+
 \varepsilon^{\mu\nu}n_\mu \tilde{\si}_\nu f^{25}
 \varepsilon^{\rho\si}n_\rho A_\si f^{26}+\tilde{\Omega}^\nu f^{27} B_\nu f^{28}+
 \varepsilon_{\mu\nu}\tilde{\Omega}^\mu f^{29} B^\nu f^{30} +
 \nonumber\\
 &+&
 n_\mu n_\nu \tilde{\Omega}^\mu f^{31} B^\nu f^{32}+
 \varepsilon^{\mu\nu}n_\mu  \tilde{\Omega}_\nu f^{33} n^\rho B_\rho f^{34}+
 n^\mu \tilde{\Omega}_\mu f^{35} \varepsilon^{\nu\rho}n_\nu B_\rho f^{36}+
 \nonumber\\
 &+&
 \varepsilon^{\mu\nu}n_\mu \tilde{\Omega}_\nu f^{37}
 \varepsilon^{\rho\si}n_\rho B_\si f^{38}+
 \tilde{\si}^\nu f^{39} B_\nu f^{40} +
 \varepsilon_{\mu\nu}\tilde{\si}^\mu f^{41} B^\nu f^{42}+
 n_\mu n_\nu \tilde{\si}^\mu f^{43} B^\nu f^{44}+
 \nonumber\\
 &+&
 \varepsilon^{\mu\nu}n_\mu  \tilde{\si}_\nu f^{45} n^\rho B_\rho f^{46}+
 n^\mu \tilde{\si}_\mu f^{47} \varepsilon^{\nu\rho}n_\nu B_\rho f^{48}+
 \varepsilon^{\mu\nu}n_\mu \tilde{\si}_\nu f^{49}
 \varepsilon^{\rho\si}n_\rho B_\si f^{50}+
 \nonumber\\
 &+&
 (\partial^\nu \tilde{\Omega}_\nu) f^{51} +
 \varepsilon_{\mu\nu}(\partial^\mu \tilde{\Omega}^\nu) f^{52} +
 n_\mu n_\nu \partial^\mu \tilde{\Omega}^\nu  f^{53} +
 \varepsilon^{\mu\nu}n_\mu \partial_\nu  (n^\rho \tilde{\Omega}_\rho)  f^{54} +
 \nonumber\\
 &+&
 n^\mu \partial_\mu (\varepsilon^{\nu\rho} n_\nu \tilde{\Omega}_\rho)  f^{55} +
 \varepsilon^{\mu\nu} n_\mu \partial_\nu (\varepsilon^{\rho\si}
 n_\rho \tilde{\Omega}_\si)  f^{56} +
 (\partial^\nu \tilde{\si}_\nu) f^{57}+
 \varepsilon_{\mu\nu} (\partial^\mu \tilde{\si}^\nu) f^{58} +
 \nonumber\\
 &+&
 n_\mu n_\nu \partial^\mu \tilde{\si}^\nu  f^{59}
 +
 \varepsilon^{\mu\nu}n_\mu \partial_\nu (n^\rho \tilde{\si}_\rho)f^{60} +
 n^\mu \partial_\mu (\varepsilon^{\nu\rho}n_\nu \tilde{\si}_\rho)f^{61} +
 \varepsilon^{\mu\nu} n_\mu \partial_\nu (\varepsilon^{\rho\si}
 n_\rho \tilde{\si}_\si)  f^{62} +
 \nonumber\\ 
 &+&
 {L}f^{63} c f^{64} +
 {\Lambda}f^{65} c f^{66} +
 {L}f^{67} \lambda f^{68} +
 {\Lambda} f^{69}\lambda f^{70}+
 n^\mu \chi_\mu L f^{71} + \varepsilon^{\mu\nu} n_\mu \chi_\nu L f^{72}+
 \nonumber\\
 &+&
 n^\mu \chi_\mu \Lambda f^{73}+\varepsilon^{\mu\nu} n_\mu \chi_\nu
 \Lambda f^{74} +
 n^\mu \chi_\mu \tilde{\si}^\nu\tilde{\si}_\nu f^{75}+
 n^\mu \chi_\mu \tilde{\si}^\nu f^{76}\tilde{\Omega}_\nu f^{77}+
 n^\mu \chi_\mu \tilde{\Omega}^\nu \tilde{\Omega}_\nu f^{78}+
 \nonumber\\
 &+&
 n^\mu \tilde{\Omega}_\mu \chi^\nu f^{79} \tilde{\si}_\nu f^{80}+
 n^\mu \tilde{\si}_\mu f^{81} \tilde{\Omega}^\nu \chi_\nu f^{82}+
 n^\mu \chi^\nu \tilde{\si}_\mu f^{83} \tilde{\si}_\nu f^{84}+
 n^\mu \chi^\nu \tilde{\Omega}_\mu f^{85} \tilde{\Omega}_\nu f^{86}+
 \nonumber\\
 &+&
 n^\mu \chi_\mu n^\nu \tilde{\si}_\nu n^\rho \tilde{\si}_\rho f^{87}+
 n^\mu \chi_\mu n^\nu\tilde{\si}_\nu f^{88}n^\rho
 \tilde{\Omega}_\rho f^{89} +
 n^\mu \chi_\mu n^\nu\tilde{\Omega}_\nu n^\rho \tilde{\Omega}_\rho f^{90}+
 \nonumber\\
 &+&
  n^\mu \chi_\mu \varepsilon^{\nu\rho}\tilde{\si}_\nu
 \tilde{\si}_\rho f^{91}+
 n^\mu \chi_\mu \varepsilon^{\nu\rho}\tilde{\si}_\nu f^{92}
 \tilde{\Omega}_\rho f^{93} +
 n^\mu \chi_\mu
 \varepsilon^{\nu\rho}\tilde{\Omega}_\nu\tilde{\Omega}_\rho f^{94} +
 \nonumber\\
 &+&
 n^\mu \tilde{\si}_\mu f^{95} \varepsilon^{\nu\rho} \chi_\nu
 \tilde{\si}_\rho f^{96} +
 n^\mu \tilde{\si}_\mu f^{97}\varepsilon^{\nu\rho}
 \tilde{\Omega}^\nu\chi_\rho f^{98}+
 n^\mu \tilde{\Omega}_\mu f^{99}\varepsilon^{\nu\rho} \chi_\nu
 \tilde{\si}_\rho f^{100}
 \nonumber\\
 &+&
 n^\mu \tilde{\Omega}_\mu f^{101}\varepsilon^{\nu\rho} \chi_\nu
 \tilde{\Omega}_\rho f^{102} +
 \varepsilon^{\mu\nu}n_\mu \chi_\nu \tilde{\si}^\rho
 \tilde{\si}_\rho f^{103}+
 \varepsilon^{\mu\nu}n_\mu \chi_\nu \tilde{\si}^\rho f^{104}
 \tilde{\Omega}_\rho f^{105}+
 \nonumber\\
 &+&
 \varepsilon^{\mu\nu}n_\mu \chi_\nu \tilde{\Omega}^\rho
 \tilde{\Omega}_\rho f^{106}+
 \varepsilon^{\mu\nu}n_\mu \tilde{\si}_\nu f^{107}\chi^\rho
 \tilde{\si}_\rho f^{108}+
 \varepsilon^{\mu\nu}n_\mu \tilde{\Omega}_\nu f^{109}\chi^\rho
 \tilde{\si}_\rho f^{110}+
 \nonumber\\
 &+&
 \varepsilon^{\mu\nu}n_\mu \tilde{\Omega}_\nu f^{111}\chi^\rho
 \tilde{\Omega}_\rho f^{112} +
 \varepsilon^{\mu\nu}n_\mu \chi_\nu \varepsilon^{\rho\si}
 \tilde{\si}_\rho \tilde{\si}_\si f^{113}+
 \varepsilon^{\mu\nu}n_\mu \chi_\nu \varepsilon^{\rho\si}
 \tilde{\si}_\rho f^{114}\tilde{\Omega}_\si f^{115}+
 \nonumber\\
 &+&
 \varepsilon^{\mu\nu}n_\mu \chi_\nu \varepsilon^{\rho\si}
 \tilde{\Omega}_\rho \tilde{\Omega}_\si f^{116} +
 \varepsilon^{\mu\nu}n_\mu\tilde{\si}_\nu  f^{117}\varepsilon^{\rho\si}
 \chi_\rho \tilde{\si}_\si f^{118}+
 \varepsilon^{\mu\nu}n_\mu \tilde{\si}_\nu f^{119}\varepsilon^{\rho\si}
 \chi_\rho\tilde{\Omega}_\si f^{120}+
 \nonumber\\
 &+&
 \varepsilon^{\mu\nu}n_\mu \tilde{\Omega}_\nu f^{121}\varepsilon^{\rho\si}
 \chi_\rho \tilde{\si}_\si f^{122}+
 \varepsilon^{\mu\nu}n_\mu \tilde{\Omega}_\nu f^{123}\varepsilon^{\rho\si}
 \chi_\rho \tilde{\Omega}_\si f^{124}+
 \nonumber\\
 &+&
 \varepsilon^{\mu\nu}n_\mu \chi_\nu
 \varepsilon^{\rho\si}n_\rho \tilde{\si}_\si f^{125}
 \varepsilon^{\omega\tau}n_\omega\tilde{\si}_\tau f^{126}+
 \varepsilon^{\mu\nu}n_\mu \chi_\mu
 \varepsilon^{\rho\si}n_\rho\tilde{\si}_\si f^{127}
 \varepsilon^{\omega\tau}n_\omega\tilde{\Omega}_\tau f^{128}+
 \nonumber\\
 &+&
 \varepsilon^{\mu\nu}n_\mu \chi_\mu
 \varepsilon^{\rho\si}n_\rho\tilde{\Omega}_\si f^{129}
 \varepsilon^{\omega\tau}n_\omega\tilde{\Omega}_\tau f^{130} +
 \varepsilon^{\mu\nu}n_\mu \chi_\nu
 \varepsilon^{\rho\si}n_\rho \tilde{\si}_\si f^{131}
 n^\omega\tilde{\si}_\omega f^{132}+
 \nonumber\\
 &+&
 \varepsilon^{\mu\nu}n_\mu \chi_\nu
 \varepsilon^{\rho\si}n_\rho\tilde{\si}_\si f^{133}
 n^\omega\tilde{\Omega}_\omega f^{134}+
 \left.
 \varepsilon^{\mu\nu}n_\mu \chi_\nu
 \varepsilon^{\rho\si}n_\rho\tilde{\Omega}_\si f^{135}
 n^\omega\tilde{\Omega}_\omega f^{137}
 \right) \ptr.
 \end{eqnarray}  
 The trivial counterterm may depend on the quantities $\xi^\mu,\ee^\mu,\zeta$ and $\eta$
 which do not appear in the total action (\ref{Sigma}). For this reason we demand the
 expression (\ref{counter}) to be invariant under the Ward-operators of the 
 vector supersymmetry and translations as well as under the Ward-operator $\cal D$. A lengthy
 and tedious analysis yields that the counterterm satisfying these conditions must vanish
 identically, so that for the determination of the trivial counterterms 
 contrary to \cite{Leitgeb} the use of the ghost equations 
 (\ref{ghostpert1}) and (\ref{ghostpert2}) is not needed any more due to the
 invariance under the symmetry $\cal D$.  
 \\[2mm]
 The last problem is devoted to the discussion of possible existance of breaking of the
 symmetries. 
 As shown in reference \cite{Leitgeb} for Landau gauge \cite{Pig} and under the assumption
 that the quantum action principle is also valid in the case of noncovariant gauges
 \cite{Buch}, the symmetries appearing in the model do not admit any anomalies and 
 are valid at the full quantum level. This completes the proof of finiteness of the model 
 to all orders of perturbation theory.

\end{document}